\documentclass[aps,pre,twocolumn,showpacs]{revtex4}
\usepackage{times}

\newcommand{\bge}{\begin{equation}}
\newcommand{\ede}{\end{equation}}
\newcommand{\ba}{\begin{array}}
\newcommand{\ea}{\end{array}}
\newcommand{\bea}{\begin{eqnarray}}
\newcommand{\eea}{\end{eqnarray}}

\newcommand{\f}{\frac}

\newcommand{\kav}{\left< k\right>}
\newcommand{\Nav}{\left< N\right>}

\newcommand{\ar}{\widetilde{\alpha}}

\usepackage{graphicx}
\usepackage{amssymb}

\begin{document}
\title{Reverse engineering of linking preferences from network restructuring}
\author{ Gergely Palla$^{\dagger\ddagger}$, Ill\'es Farkas$^{\dagger}$,
 Imre Der\'enyi$^{\ddagger}$, Albert-L\'aszl\'o Barab\'asi$^{\nmid}$ 
and Tam\'as Vicsek$^{\dagger\ddagger}$}
\affiliation{$^{\dagger}$Biological Physics Research Group of HAS, P\'azm\'any P.\ stny.\ 1A, H-1117 Budapest, Hungary, \\ 
$^{\ddagger}$ Dept. of Biological Physics, E\"otv\"os University,
 P\'azm\'any P.\ stny.\ 1A, H-1117 Budapest, Hungary, \\
$^{\nmid}$Dept. of Physics, University of Notre Dame, IN 46566, USA}

\begin{abstract}
We provide a method to deduce the preferences governing
 the restructuring dynamics of a network from the observed
 rewiring of the edges. Our approach is applicable for systems in which the 
preferences can be formulated in terms of a single-vertex energy function
with $f(k)$ being the contribution of a node of degree $k$  
to the total energy, and
 the dynamics obeys the detailed balance. The method is
 first tested by Monte-Carlo simulations of restructuring graphs with
 known energies, then it is used to study variations of real network 
systems ranging from 
the co-authorship network of scientific publications to the asset
 graphs of the New York Stock Exchange. The empirical energies
obtained from the {\it restructuring} can be described by a universal
 function $f(k)\sim -k\ln k$, which is consistent with and justifies 
the validity of the preferential attachment rule 
proposed for {\it growing} networks.  
\end{abstract}
\maketitle

\section{Introduction}

In the past few years, the analysis of the network structure of
 complex systems has become a rapidly expanding interdisciplinary field 
\cite{b-a-revmod,dorog-mendes-book,Vespignani-book}.
Interpreting the dynamics of these complex networks
 with techniques developed in statistical mechanics represents 
a new approach to network theory
\cite{burda,burda2,berg-laessig,manna-hamiltonian,dms-principles,derenyi,
palla,farkasspringer}. A recently introduced method is
 to define equilibrium network ensembles as stationary
ensembles of graphs generated by restructuring processes obeying
detailed balance
and ergodicity \cite{farkasspringer}.
During such a restructuring process, edges are
removed and/or inserted. If an {\it energy function} of the graphs is
 also given, then the construction of  the corresponding microcanonical,
canonical and grand canonical ensembles can be carried out 
in a similar way
 to classical statistical physics.
 However, unlike in many
physical systems, the energy of a graph 
cannot be derived from first principles and does not necessarily exist.
 In case of  graph ensembles
 designed to have some prescribed property, 
a properly defined  cost function (or energy) can capture the deviations from
 this property. 
This method of defining energy is typically used in optimization contexts.

For network systems, in which the restructuring is assumed to be
 governed by interactions that can be interpreted in terms
 of energy, one can try to 
deduce this energy using the data obtained from the time series of the 
evolution of the system.
In the present work, our goal is to demonstrate this 
{\it reverse engineering} process in detail.
In particular, we shall put our methods to practice for a few different
 network systems such as the co-authorship networks 
 of scientific publications 
\cite{newman_collab1,newman_collab2,newman_collab3,barab_collab,jeong_paper,simeon_paper},
 the network of the 
 US film actors \cite{watts_actors,barab_paper}, and the asset graphs of the 
 New York Stock Exchange \cite{Kertesz}.

\section{Restructuring processes and preferences}
\label{restruct}
In the following we assume that the preference of establishing a given
 link can be expressed in terms of a change in the network's total energy.
During this graph restructuring process, the edges (which can be deleted, 
relocated or newly inserted) can be thought of
 as particles in a physical system and the vertices as the volume of the
 system. Since the properties of growing networks have already been
 studied in great details, here we assume that the number of vertices,
$N$, is fixed.


  In the statistical mechanics picture, the time evolution of the 
probability of occurrence of the graphs
is governed by a set of master equations \cite{farkasspringer}:
\bea
\f{\partial P_a}{\partial t} = \
\sum_b ( P_b r_{b\to a} - P_a r_{a\to b} ) \, ,
\eea
where $P_a$ is the probability of graph $a$ and $r_{a\to b}$ denotes
the transition rate from graph $a$ to graph $b$.
If the dynamics is governed by an energy function, then it determines the 
ratio of the 
transition rates as
\bea
\f{r_{a\to b}}{r_{b\to a}}=e^{-(\Delta E_{ab}-\mu \Delta M_{ab})/T}.
\label{eq:rates}
\eea
In the equation above $\Delta E_{ab}=E_b-E_a$ is the energy difference 
between the graphs, $M_{ab}=M_b-M_a$ denotes the difference in the total number 
of edges, $\mu$ is the chemical potential associated with the
 appearance of extra edges in the system
 and $T$ denotes the temperature, corresponding
 to the level of noise.
If $M$ is kept constant, then in the $T\to\infty$ limit 
the dynamics converges to a totally random rewiring
process, and thus, the ensemble of classical random graphs \cite{ErdosRenyi} 
is recovered.
 On the other hand, at low temperatures the
graphs with lowest energy (meeting the requirements set up by the preferences)
 are selected with enhanced probability.

As the transition rates obey detailed 
 balance if the dynamics is ergodic,
the probability distribution
 will converge to a stationary distribution.

\section{Single-vertex energies}
\label{Singlevertenergy}
The energy function of a network corresponds to the preferences
 in the rearranging of the graph structure (e.g. when relocating
 an edge from one vertex to another). if only the local properties
 of the graph influence the edge relocation, then the simplest form of
 the energy is given by the sum of contributions from the individual vertices.
 Assuming 
that these contributions depend only on the degrees of the vertices,
the total energy can be written as
\bea
E=\sum_{i=1}^{N}f(k_i), \label{eq:f-sforma}
\eea
where $N$ is the total number of vertices, and $k_i$ denotes the degree
 of vertex $i$. 
Note that if $f(k)$ 
is shifted by a constant, the resulting dynamics
 remains unchanged since  only the {\it difference between energies}
associated with different degrees may influence the restructuring.
Therefore, without loss of generality we may set $f(0)=0$.
 Furthermore, 
the linear part of $f$ is irrelevant
if the number of edges is constant
(since its contribution is 
proportional to the number of edges in the graph),
and simply renormalizes the chemical potential
if the number of edges is allowed to change. 

 An alternative form of 
the single-vertex energy functions can be written as
\bea
E=\sum_{i=1}^{N}\sum_{i'}g(k_{i'}), \label{eq:g-sforma}
\eea
where $i'$ runs over all vertices that are neighbors of vertex $i$.
In this interpretation, the energy of an individual vertex
 depends on the connectivities of its neighbors, and 
 vertex $i$ collects an energy $g(k_{i'})$ from each of its neighbors. 
These neighbors in turn will all collect $g(k_i)$ from vertex $i$, 
therefore, the total contribution to the energy from vertex $i$ is 
$k_i g(k_i)$. Thus, by choosing
\bea
f(k)&=&kg(k),\label{eq:fesg}
\eea
 the two alternative forms of the single-vertex
 energy, (\ref{eq:f-sforma}) and (\ref{eq:g-sforma}), become equivalent.

The advantage of the latter representation is that the irrelevant
linear part of $f(k)$ appears as a simple additional constant term in  $g(k)$ 
 for $k\geq1$. 

\section{Reverse engineering with independent edge dynamics: the method}
In the following we assume that the energy
function of the investigated system falls into the class
 of the single-vertex energy functions (\ref{eq:g-sforma}).
 The main point of our method is that
it assumes {\it independent dynamics for 
 each half edge}.
 In this approach, each vertex $i$ is treated as a 
hedgehog, with $k_i$ half edges. Since the energy function depends
 only on the degree sequence, the energy of the network
is identical to that of the set of isolated hedgehogs.
The restructuring of the network can be approximated by the 
 insertion and deletion of half edges.
 This approximation
 is valid when the number of vertices is large compared to the typical
 number of half edges on the individual vertices.
Let us denote by  $r_{k\rightarrow k-1}$ the rate at which existing half 
edges on vertices with degree $k$ disappear, and accordingly by  
 $r_{k-1\rightarrow k}$ the rate at which new half edges appear on 
vertices with degree $k-1$. The ratio of these two rates is determined
 by the energy in a way similar to Eq. (\ref{eq:rates}):
\bea
\f{r_{k\rightarrow k-1}}{r_{k-1\rightarrow k}}=\f{e^{(f(k)-\tilde{\mu} k)/T}}
{e^{(f(k-1)-\tilde{\mu}(k-1))/T}}, \label{eq:ratio}
\eea
where $\tilde{\mu}=\mu/2$ denotes the chemical potential for the 
half edges. From (\ref{eq:ratio}) it is clear that the effect of the
 chemical potential
 on the dynamics cannot be separated from 
 the effect of the linear part of $f(k)$, and the multiplication
of $f(k)$ and $\mu$ by the same number is equivalent to the rescaling
 of the temperature. For these reasons it is convenient to introduce
the dimensionless single-vertex energy
\bea 
\phi(k)=\f{f(k)-\tilde{\mu} k}{T}, \label{eq:phi(k)}
\eea
which incorporates the chemical potential as well.
This yields
\bea
\f{r_{k\rightarrow k-1}}{r_{k-1\rightarrow k}}=e^{\phi(k)-\phi(k-1)}.
\label{eq:phi_ratio}
\eea

\begin{figure}[t!]
\centerline{\includegraphics[width=0.8\columnwidth]{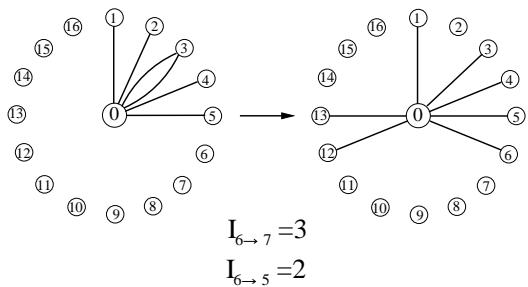}}
\caption{We illustrate the measurement of the currents
 $I_{k\rightarrow k+1}$ and
$I_{k\rightarrow k-1}$ in the dataset by comparing two subsequent states of
 vertex $0$ in a
 hypothetical simple graph with $17$ vertices. 
(Only edges connected to vertex $0$
 are shown). In the
 initial state (left side), the central vertex is connected to  
vertices $1,2,4,5$ with one edge and to vertex $3$ with two edges, hence
 its degree is $k=6$.  In the second state (right side), 
the link to vertex $2$ and one of the links to vertex $3$ 
disappeared giving a 
 contribution $I_{6\rightarrow5}=2$ to the currents. Similarly, 
new links to vertices $6,12$ and $13$ appeared, thus
$I_{6\rightarrow7}=3$. }
\label{fig:ind_edges}
\end{figure}

The number of disappearing edges on vertices with degree $k$ per unit time 
can be written as
\bea
I_{k\rightarrow k-1}=r_{k\rightarrow k-1}p_kNk,
\label{eq:I_k->k-1}
\eea
because there are $p_kN$ vertices with $k$ edges and these have a total of
 $p_kNk$ edges on them. Similarly,
 the current between the same two states in the opposite direction is
\bea
I_{k-1\rightarrow k}=r_{k-1\rightarrow k}p_{k-1}N(N-1),
\label{eq:I_k-1->k:multi}
\eea
because there are $p_{k-1}N$ vertices with $k-1$ edges and the appearing new 
half edge can be placed at $N-1$ different positions (corresponding to the 
$N-1$ possible neighbors of a given vertex). For simple graphs (i.e., when
 only zero or one connection is allowed between two vertices) the number
 of these possible positions is $N-1-(k-1)$ and accordingly the $(N-1)$ factor
 in expression (\ref{eq:I_k-1->k:multi}) should be replaced by $(N-k)$.
 However,
 in all cases studied in this paper we have $N\gg k$, and therefore we may use
\bea
I_{k-1\rightarrow k}=r_{k-1\rightarrow k}p_{k-1}N^2,
\label{eq:I_k-1->k:simple}
\eea
in all situations.
From the ratio of the currents (\ref{eq:I_k-1->k:simple}) and 
(\ref{eq:I_k->k-1}) and Eq. (\ref{eq:phi_ratio}) we get
\bea
e^{\phi(k)-\phi(k-1)}=\f{I_{k\rightarrow k-1}}{I_{k-1\rightarrow k}}
\f{p_{k-1}}{p_k}\f{N}{k}.
\label{eq:expbetaf:multi}
\eea
The quantities appearing on the right hand side can be
 obtained from the available data and
the currents can be computed as follows.
 At each time step each vertex should be compared to its own state
 in the next time step. Those links that
appear only in the latter graph contribute a current of amount 1 to 
$I_{k\to k+1}$, where
 $k$ is the degree of the vertex in the former state, whereas those links
 that are deleted in the time step  contribute to $I_{k\to k-1}$.
If a new vertex appears in the second graph, each of its links will
 contribute to $I_{0\to 1}$ with a current of amount 1.
 Similarly, links of disappearing vertices of degree $k$ will
 contribute to $I_{k\to k-1}$.
 A simple example demonstrating the evaluation
of the currents in the frame of independent edge dynamics is shown in
 Fig.\ \ref{fig:ind_edges}.

Since we set $f(0)=0$, the corresponding $\phi(k)$ will also be zero for
$k=0$. For $k\geq 1$, we can express $\phi(k)$ using the recursion relation
 (\ref{eq:expbetaf:multi}) as
\bea
\phi(k)=\sum_{l=1}^k\ln\left[\f{I_{l\rightarrow l-1}}{I_{l-1\rightarrow l}}
\f{p_{l-1}}{p_l}\f{N}{l}\right]. 
\label{eq:f_k_emp}
\eea
In the  alternative representation of the energy defined 
 in equations (\ref{eq:g-sforma}) and (\ref{eq:fesg}), one can introduce
\bea
\gamma(k)=\f{\phi(k)}{k}=\f{g(k)-\tilde{\mu}}{T}.
\eea


\section{Case Studies} 

We first tested our method on known energy functions
used in Monte Carlo simulations, then applied it to study real world systems
 ranging from the co-authorship
 networks of scientific publications to the asset graphs of the
 New York Stock Exchange. In each case, the statistics of vertices with a given degree became
 poor with increasing $k$, therefore only the interval of the well represented 
degrees was taken into account in our reverse engineering process.
 The upper boundary of this interval, $k_{\rm max}$,
was set to the point where the time average of the number of vertices 
with degree $k$ became
 smaller than 0.1\% of the average number of vertices.

As explained in the previous sections, when the total number 
 of edges, $M$, is fixed, the linear part of $f(k)$ is 
irrelevant. When $M$ is allowed to vary, a linear function, $-\tilde{\mu} k$,
naturally appears in Eq. (\ref{eq:rates}), and cannot be uniquely separated
 from the energy. Therefore, we will focus
 on the non-linear part of the empirically
 obtained $\phi(k)$, corresponding to the non-linear part of the single-vertex
 energy. In general, the $\phi(k)$ obtained from the data 
can be  decomposed into
\bea
\phi(k)=C_0+C_1k+\phi^*(k),\label{eq:fit_phi_k}
\eea
where $\phi^*(k)$ denotes the non-linear part of $\phi(k)$.
 The additional
 $C_0$ constant shift is introduced to account for the irregular
 behavior of the energy of vertices with zero degrees in the collaboration
 networks (the networks of
 the scientific publications and the US movie actors).
 In these systems $\phi(k)$ is relatively smooth for
 $k\geq 1$ in contrast to a
 gap in the energy observed between $k=0$ and $k=1$.
 Accordingly, $k=0$ has been excluded
 when fitting the empirical results, and the shift of the $k\geq 1$
 part caused by the gap is taken into
 account with the help of $C_0$. The $\gamma(k)$ corresponding to
(\ref{eq:fit_phi_k}) can be written as
\bea
\gamma(k)=C_0/k+C_1+\gamma^*(k), \label{eq:fit_gamma_k}
\eea
 where $\gamma^*(k)=\phi^*(k)/k$. In the expression above, the
 first term decays for high $k$ and the second is a constant, therefore,
 in practical cases the remaining non-linear part is somewhat more
 apparent than in the $\phi(k)$ representation.

 In the figures showing our results, we shall plot $\phi(k)-C_1k$
and $\gamma(k)-C_0/k-C_1$, since this way both the non-linear
 part of the energy and the gap at $k=0$ for the collaboration networks 
become  apparent.

\subsection{Testing with known energy functions}

We first tested our method on known energy functions used in
 Monte Carlo simulations with a constant ($N=10000$) number of vertices and 
$\kav=2$.
The most trivial choice for the energy function was the case when no
 energy was present in
 the rewiring and the edges were replaced randomly. In this case 
 the network converges to a classical random graph regardless of
 the initial state, and accordingly the maximum degree taken into account in
 the reverse engineering process
 was  $k_{\rm max}=7$. The other 
 tested energy was chosen to be $f(k)=-k\ln(k)$. This choice 
was motivated by our previous studies which
showed that  in this 
 system, around $T=1$ a scale-free degree distribution can be observed
\cite{palla}, hence the interval of $k$ taken into account in the reverse
 engineering is wider ($k_{\rm max}=27$) than in case of the random rewiring.

\begin{figure}[t!]
\centerline{\includegraphics[angle=0,width=1.0\columnwidth]{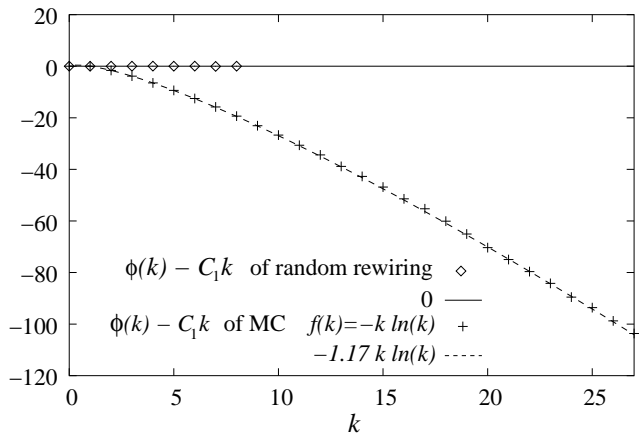}}
\caption{$\phi(k)$ obtained from the restructuring data
 of the 
 MC simulation with $C_1k$ subtracted, together with fits. 
For the case of $E=0$, (random rewiring, circles), 
the system converges to a classical
 random graph, which results in a narrow degree distribution.
 In the system with $E=-k\ln(k)$ (crosses), the interval in which $\phi(k)$
 can be obtained is wider. 
Thus the results from reverse engineering are in good agreement with
 the initial MC energies.}
\label{fig:teszt_fig}
\end{figure}

Fig. \ref{fig:teszt_fig}. shows our results for the two test energies.
In both cases, $C_0$ 
 was set to zero, and $C_1$ was set equal to $\phi(k=1)$, (since both energies
 used in the MC simulations
 give $f(1)=0$).
 In both situations, the resulting empirical $\phi^*(k)$ could be
 fitted well with the energy used in the MC simulation.

\subsection{Collaboration networks}

The collaboration networks investigated in this paper can be listed
 as follows: the co-authorship networks of neuroscience and math 
publications, the co-authorship networks
 of the Los Alamos e-print archives of Condensed Matter, 
High Energy Physics - Phenomenology and Astrophysics,
 and the 
 network of the US movie actors.
 These networks are all  bipartite: the 
collaborators (scientists, actors) and the collaborations 
(publications, movies) can be represented by two different kinds of vertices,
and the links in the system are always placed between vertices of 
 different types (a scientist is connected to the papers in which he/she is
 a coauthor). Uni-partite projections of these graphs can be made by keeping 
only one
 kind of the vertices and placing a link between them if they are
 linked to a common vertex of the other type in the bipartite graph.
In the cases studied in this paper, the projection 
 onto the {\it graphs of collaborators} has been used: the nodes 
 represent collaborators (scientists or movie actors) and two nodes
 are linked if they have a common collaboration (articles, movies).
 In the graphs obtained this way, both edges and vertices are distinguishable
 and multiple links between vertices are allowed.

The collaboration network databases we have used 
\cite{database,simeon_data} provide
 the time of the emergence of the links, 
but the times when they disappear are not defined.
In other words, the time we consider an edge to be alive 
 after its birth influences the structure of the graphs representing
 the system year after year.
 This parameter will be
 denoted by $\tau$, the {\it life-time of an edge}. The minimal value
 for $\tau$ is
 equal to the time resolution of the dataset. In case 
 of $\tau=\infty$, the system is simply growing and no edges are deleted.
 Evidently, the choice of $\tau$ may influence all statistical properties
 of the network. 

The number of vertices in these systems is changing slowly with time, but 
the number of edge restructuring events per time step is larger by two
 orders of magnitude than the change in the number of vertices. 
 Therefore, the effect of the increase or decrease in $N$ can be neglected,
and we use the time average of the right hand side of Eq.
(\ref{eq:expbetaf:multi}) (derived
 for a fixed $N$) to calculate $\phi(k)$.

 The most important statistical properties of the studied
 networks are summarized in 
 Table \ref{tab:collab}.\\

\begin{table*}[hbt!]
\begin{tabular}{|l|c|c||c|c|c|c||c|c|c|c||c|c|c|c|}\hline
                &        &   & \multicolumn{4}{c||}{$\tau=1$ year} & \multicolumn {4}{c||}{$\tau=2$ year} & \multicolumn{4}{c|}{$\tau=3$ year} \\ \hline
                & timespan & timestep & $\Nav$ & $\kav$ & $k_{\rm max}$ & $\ar$ &$\Nav$ & $\kav$ & $k_{\rm max}$ & $\ar$ & $\Nav$ & $\kav$ & $k_{\rm max}$ & $\ar$ \\ \hline
 movie actors     &1896-1999 & 1 year & 9223 & 28.04&  100 &0.56 &15350 &34.52 & 130 &0.69 & 20373 & 39.23&  150 & 0.74  \\ \hline \hline
neurosci. publ. &1991-1998 & 1 year & 55057 &7.837 & 35 &0.86 &87513 &9.397  &  38 &1.02 &113160 & 10.54 & 44 &0.96    \\ \hline
math publ.       &1991-1999 & 1 year & 17279 & 2.737& 15 &0.83 &29116  & 3.141 & 17 &0.93 &38633 & 3.45 & 20 & 0.99    \\ \hline \hline 
hep-ph           &'92.03-'04.02 & 1 month &  3202  & 7.14  & 23&0.41 &4386  & 8.52   & 30 &0.58 &5186 & 8.48 & 35 & 0.62 \\ \hline
astro-ph         & '92.04-'04.02& 1 month & 5472   & 10.90   & 60 &0.44 &7103  & 13.02    & 70 &0.56 & 7891 & 13.98 & 80 & 0.59 \\ \hline
cond-mat         &'92.04-'04.02& 1 month & 5648 & 4.16 & 29 & 0.73 & 7464 & 4.71 & 34 &0.96 & 8194 & 4.94 & 37 &0.91\\ \hline    
\end{tabular}
\caption{The most important statistical properties of the studied collaboration
 networks.  $k_{\rm max}$ denotes the maximal degree which occurred
 frequently enough in the restructuring process to be able to
 extract the energy, and $\ar$ is the fitting parameter introduced in Eq. (\ref{eq:phi_fit}).
\label{tab:collab}}
\end{table*} 

\subsubsection{Neuroscience- and math publications }
\begin{figure}[t!]
\centerline{\includegraphics[angle=0,width=1.0\columnwidth]{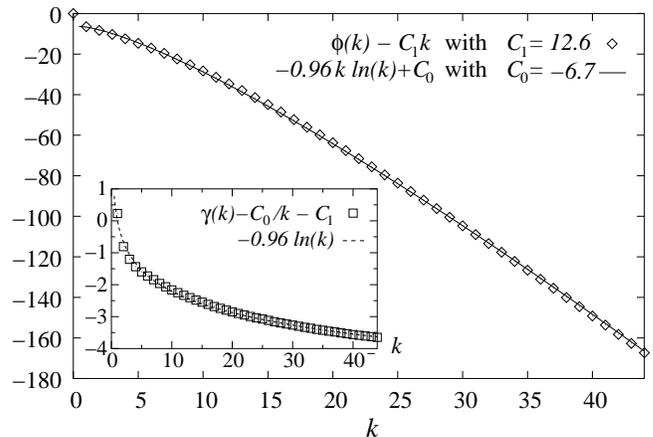}}
\caption{ The empirical $\phi(k)$ obtained for the neuroscience
 publication network (circles) for $\tau=3$. In order to make the non-linear
 part of $\phi(k)$ more apparent, the linear part of the fit, $C_1k$ was
 subtracted from the data. 
The actual values obtained for the fitting parameters 
 were $C_0=-6.7,C_1=12.6,\ar=0.96$. A relatively large gap between 
$\phi(k)$ at $k=0$ and $k=1$ can be observed. 
 The inset shows the corresponding $\gamma(k)$ with $C_0/k+C_1$ subtracted
 from the data (boxes) together with
 $-\ar\ln(k)$ (dashed line).}
\label{fig:neuro}
\end{figure}

Data on these two systems was obtained from Ref. \cite{database}.
The time resolution of the appearance of the new articles was 
one year in the dataset. The upper boundary of the interval of $k$ taken 
into account in the reverse engineering was in the range of $20-40$. 
The results in case of the neuroscience papers for $\tau=3$ years are 
shown in Fig. 
\ref{fig:neuro}, and very similar plots
 were obtained for the math publications. 
In both cases, $\phi^*(k)$ and $\gamma^*(k)$ can be 
fitted with 
\bea
\phi^*(k)=-\ar k\ln(k), \qquad \gamma^*(k)=-\ar \ln(k) \label{eq:phi_fit}
\eea
where $\ar$ is a positive constant. The actual value of this constant
was $\ar=0.96$ for the neuroscience publications and $\ar=0.99$ in case of
 the math papers.

 In both cases, $\phi(k=0)$ is separated from the $k\geq 1$ part
 by a gap, indicating that zero connection is unfavorable for the vertices.
This is consistent with the observed degree distributions, in
 which the vertices of zero degree were suppressed. (In case of
  the math papers, vertices with zero degree were totally absent).

For $\tau=1$ and $\tau=2$, similar results can be obtained with
 modified fitting parameters.\\

\subsubsection{The Los Alamos e-print archives}

We also studied the restructuring of the co-authorship networks of 
 three e-print archives:  Condensed Matter, 
High Energy Physics - Phenomenology and Astrophysics \cite{simeon_data}.
In all three cases, the time resolution of the appearance
 of the new manuscripts was one month. To be able
 to compare the properties of these systems to the previously investigated
 co-authorship networks, we set $\tau=1,2,3$ years in these 
cases as well. From the point of view of the energy,the obtained 
behavior was very similar to the previous cases.
For the  Condensed Matter, 
High Energy Physics - Phenomenology networks,  $k_{\rm max}$ was in the range
 of that seen for
the neuroscience publications, and
 for the Astrophysics archive it was somewhat larger ($k_{\rm max}=80$).
The non-linear part of the empirical energies could be fitted with functions 
of the form given in (\ref{eq:phi_fit}). The value
 of the parameter $\ar$ in case of $\tau=3$ years was 
$\ar=1.05,\ar=0.77,\ar=0.59$ for
 the Condensed Matter, High Energy Physics - Phenomenology and Astrophysics
 archives respectively. 

\begin{figure}[t!]
\centerline{\includegraphics[angle=0,width=1.0\columnwidth]{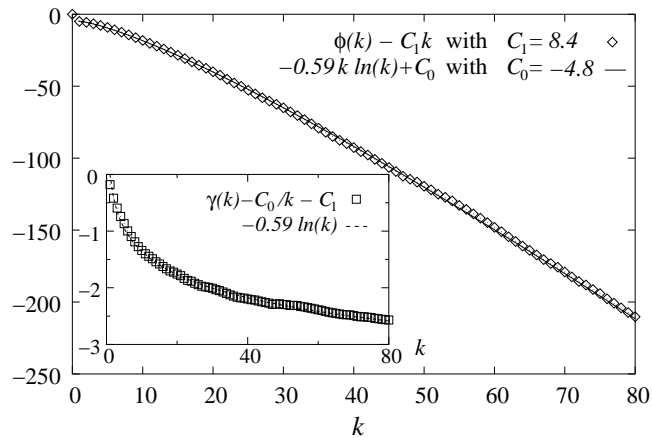}}
\caption{ The empirical $\phi(k)$ of the Astrophysics archive for $\tau=3$ years
 with $C_1k$ subtracted (circles) plotted together
 with $-\ar k\ln(k)+C_0$ (solid line). The values of the fitting parameters read 
$C_0=-4.8,C_1=8.4,\ar =0.59$. The $\phi(k=0)$ is separated from the $k\geq1$ part 
by a gap, similarly to the case shown in 
Fig.\ref{fig:neuro}. In the
 inset  the corresponding $\gamma(k)-C_0/k-C_1$ is shown together 
 with $-0.59\ln(k)$.}
\label{fig:astro}
\end{figure}

 The
 results for the Astrophysics archive are shown in Fig.\ref{fig:astro}.
 The gap at $k=0$ in the energy together with
 the suppression of the vertices with zero degrees seen in case
 of the neuroscience- and math publications is also present
 in all three systems. \\

\subsubsection{The network of movie actors}

\begin{figure}[t!]
\centerline{\includegraphics[angle=0,width=1.0\columnwidth]{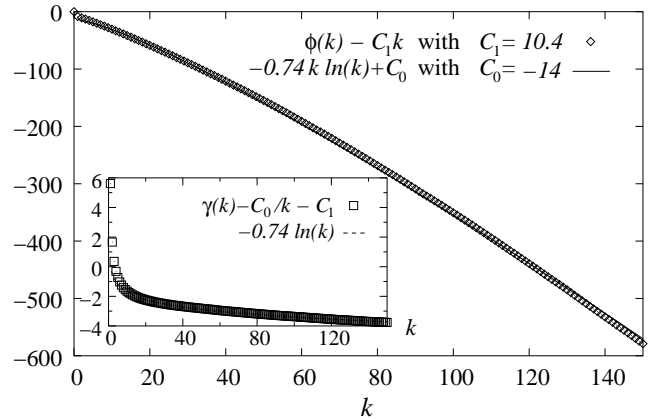}}
\caption{ $\phi(k)$ obtained in case of the US movie actors network
 for $\tau=3$ with $C_1k$ subtracted from the data (circles). The resulting
 functions can be fitted with $-\ar k\ln(k)+C_0$ similarly to previously 
studied collaboration networks. The actual values of the fitting
 parameters are $C_0=-14,C_1=10.4,\ar =0.74$. The gap separating $\phi(k=0)$ from
 the $k\geq1$ part, characteristic for the other collaboration networks
 can be seen in this case as well. The 
$\gamma(k)$ corresponding to this $\phi(k)$ is shown in the inset with 
$C_0/k+C_1$ subtracted from the data, plotted together
 with $-\ar \ln(k)$.  }
\label{fig:actors}
\end{figure}

In the data of the US movie actor network \cite{database},
the time resolution of the appearance of new films was 
one year.
For this network, the allowed $k$ interval in the
 reverse engineering is reasonably wide. Similarly to the co-authorship 
networks studied in the previous paragraphs, the non-linear part of
 the empirical energy can be fitted with functions of the form
 given in (\ref{eq:phi_fit}), and a gap at $k=0$ together
 with the suppression of the 
 vertices with zero degrees can be observed. Our results
 for $\tau=3$ years are shown in Fig.\ref{fig:actors} with $\ar =0.74$.

\subsection{Asset graphs of the New York Stock Exchange}

\begin{figure}[t!]
\centerline{\includegraphics[angle=0,width=1.0\columnwidth]{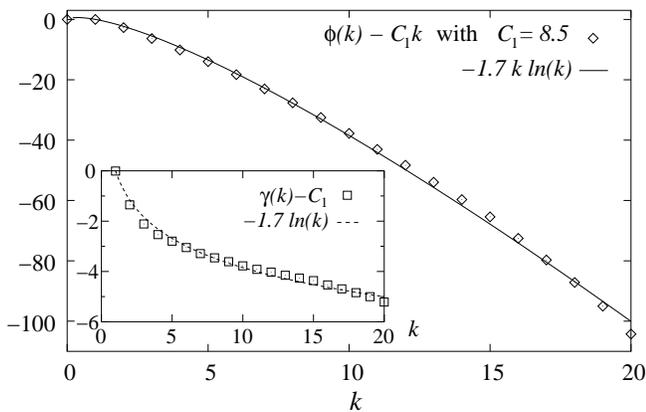}}
\caption{ The empirical $\phi(k)$ of the asset graphs of  
the New York Stock Exchange with $C_1k$ subtracted from the data (circles),
 plotted together with $-\ar k\ln(k)$ (solid line). The fitting parameters
 were $C_0=0,C_1=8.5,\ar =1.7$. The inset shows
 $\gamma(k)-C_1$ plotted together with $-\ar \ln(k)$.}
\label{fig:stock}
\end{figure}

In this study, we used the asset graph sequence received from 
\cite{asset_data}.
These graphs were constructed from
 correlations of the 477 NYSE traded stocks: in a given time window
 the 476 most strongly correlated pairs of papers were linked \cite{Kertesz}.
 Therefore in this case the obtained graphs were simple graphs. 
The market data was recorded with a time window width of 1000
days, from 02-Jan-1980 to 31-Dec-1999, and the window was stepped monthly. This resulted in
 195 time windows. In this case, the general rule used to
 determine $k_{\rm max}$ would result in accepting such degrees also
 that are represented by less than one vertex on average,
 therefore $k_{\rm max}$ was set equal to the highest degree 
 represented by at least one vertex on average, yielding $k_{\rm max}=20$. 

Similarly to the collaboration networks, the non-linear part of the 
 empirical energy can be fitted using (\ref{eq:phi_fit}) with
 $\ar =1.7$, as shown in Fig. \ref{fig:stock}. For this
 system, the gap in the energy at $k=0$ seen in the case of the collaboration
 networks is absent, and accordingly the $C_0$ fitting parameter can be set
 to zero.

\section{Discussion}
For each studied natural network, the non-linear part of the empirical
 single-vertex energy can be fitted with 
 the same universal function given in Eq. (\ref{eq:phi_fit}). We note that 
(due to the 
 relatively narrow intervals in $k$) power-laws of the
 form $E=-B\sum_ik_i^{\beta}$ can be used for fitting equally well,
 with fitting parameters $B$ and $\beta$. However, the following 
 two reasons make the logarithmic fit,  $E=-\ar \sum_ik_i\ln(k_i)$, 
far more favorable:
 first, this latter has fewer 
 fitting parameters; and second, since in each example $\ar$ has been found to
 be very close to unity (within a factor of two), even $\ar$ is
 not a real fitting parameter.

The logarithmic fit is also consistent with  the rule
 of preferential attachment observed in growing
 networks \cite{jeong_paper,newman_pref,redner_pref}.
 This can be shown by estimating the energy
 change of a vertex by the derivative of  $\phi^*(k)+\ar k=-\ar k(\ln(k)-1)$, 
yielding $\Delta E/T=-\ar \ln(k)$ (the irrelevant linear term $\ar k$ has 
 been added to $\phi^*(k)$ for simplicity).
Plugging this into the Boltzmann factor, $\exp[-\Delta E/T]$,
the resulting acception/rejection ratio
for a randomly selected move is proportional to $k^{\ar}$. In case of 
$\ar=1$ we
 recover the linear preferential attachment rule. For several growing
 networks, however, 
 the preferential attachment was found to scale as $k^{\alpha}$
 \cite{jeong_paper,redner_pref}.
 The values for $\alpha$ measured by Jeong et al.\cite{jeong_paper} 
 are close to those we obtained for $\ar$. Hence our results provide
 a novel justification of the preferential attachment rule.

 Furthermore, 
in  the second representation of the energy, in which vertices contribute to
 the
 total energy depending on the degrees of their neighbors, the 
energy contribution of vertex $i$ becomes 
$\sum \gamma^*(k_{i'})=-\ar \sum_{i'}\ln(k_{i'})$, where
 $i'$ runs over the neighbors of vertex $i$. 
This can be interpreted as follows: the vertices ``feel'' advantageous
  to be linked to other vertices of high degree, hence the magnitude
of a neighbor's contribution to the energy of vertex $i$ is 
 a monotonously increasing function of that neighbor's degree.
The distinction between the neighbors in this picture follows
 the general Weber-Fechner law of sensation:
 a stimulus  of a given degree generates a
 perception proportional to the logarithm of that degree in the neighbors.
This logarithmic law of perception is natural in cases where the 
intensity of the stimulus may vary over several orders of magnitude, which
 is the case for degrees in scale-free networks.
  
Another situation where the use of $g(k)$ is very convenient
 is when it is advantageous to
 reach many other vertices via short paths, but it is 
expensive to have lots of connections. 
Under such circumstances the best strategy
 is to connect to a few other vertices with high degrees: numerous vertices
 can be reached in two steps at the cost of few own links. If the 
 number of total edges is allowed to vary, such a scenario could simply be 
modeled by a decreasing $g(k)$ function (such as $-\ln (k)$) 
 together
 with a negative chemical potential (corresponding to a linear penalty 
 for the own links of a vertex).

\section{Summary}
We developed a reverse engineering method to deduce
 the preferences governing the restructuring in non-growing networks
 from the statistics of the observed relocations. Our approach
 is applicable to systems where the preferences can be 
interpreted by single-vertex energy functions and the 
 dynamics obeys detailed balance. The method 
 was first tested on Monte-Carlo simulations ran with known energy functions
 yielding reassuring results. Real networks such as 
the co-authorship network
 of scientific publications, the network of the US movie actors
 and the asset graphs of the New York
 Stock Exchange were also studied.

 In each case, the non-linear
 part of the single-vertex energy function could be fitted with 
 the same universal function $E=-\ar \sum_i k_i\ln(k_i)$ with 
$0.5<\ar <2$. This energy is shown to be consistent with the preferential 
attachment rule of growing networks.

 Furthermore, the obtained energy
 can be alternatively written as $E=-\ar \sum_i\sum_{i'}\ln(k_{i'})$, where
 the summation for $i'$ runs over the neighbors of vertex $i$. In this
 representation the energy contributions of the vertices can be
interpreted by the general Weber-Fechner law of sensation:
the stimulus of a given vertex generates a perception in its
 neighbor proportional
 to the logarithm of that degree.

 


\acknowledgments
The authors are grateful to Hawong Jeong for providing the 
database of the US movie actors, the neuroscience and
 math publications; to Simeon Warner for providing the  
 dataset of the e-print archives; and to Jukka-Pekka Onnela
 and J\'anos Kert\'esz for providing the asset graph sequence
 of the NYSE. This work was supported in part by the 
 Hungarian Science Foundation OTKA Grants No. F047203 and T034995.


\begin{thebibliography}{99}
%
\bibitem{b-a-revmod}
A.-L. Barab\'asi and R. Albert,
Rev.\ Mod.\ Phys.\ {\bf 74}, 47 (2002).
\bibitem{dorog-mendes-book}
S. N. Dorogovtsev and J. F. F. Mendes,
{\em Evolution of Networks: From Biological Nets to the Internet and WWW}
(Oxford University Press, 2003).
\bibitem{Vespignani-book}
R. Pastor-Satorras, A. Vespignani,
{\em Evolution and Structure of the Internet}
(Cambridge University Press, 2004).
%
%
\bibitem{burda}
Z. Burda, J. D. Correia, and A. Krzywicki,
Phys.\ Rev.\ E {\bf 64}, 046118 (2001).
\bibitem{burda2}
Z. Burda and A. Krzywicki,
Phys.\ Rev.\ E {\bf 67}, 046118 (2003).
\bibitem{berg-laessig}
J. Berg and M. L{\"a}ssig,
Phys.\ Rev. Lett.\ \textbf{89}, 228701 (2002).
\bibitem{manna-hamiltonian}
M. Baiesi and S. S. Manna,
Phys.\ Rev.\ E\ \textbf{68}, 047103 (2003).
\bibitem{dms-principles}
S. N. Dorogovtsev, J. F. F. Mendes, and A. N. Samukhin,
Nucl.\ Phys.\ B\  \textbf{666}, 396 (2003).
\bibitem{derenyi}
I. Der{\'e}nyi, I. Farkas, G. Palla, and T. Vicsek,
Physica A  {\bf 334}, 583-590 (2004).
\bibitem{palla}
G. Palla, I. Farkas, I Der{\'e}nyi, and T. Vicsek,
Phys.\ Rev.\ E   {\bf 69}, 046117 (2004).
\bibitem{farkasspringer}
I. Farkas, I. Der{\'e}nyi, G. Palla, and T. Vicsek,
{\tt cond-mat/0401640} (to appear in Lecture Notes in Physics).
%
%
\bibitem{newman_collab1}
M. E. J. Newman,
Proc. Natl. Acad. Sci. USA 98, 404-409 (2001).
\bibitem{newman_collab2}
 M. E. J. Newman,
Phys.Rev. E {\bf 64} 016131 (2001).
\bibitem{newman_collab3}
 M. E. J. Newman,
 Phys.Rev. E {\bf 64} 016132 (2001).
\bibitem{barab_collab}
 A.L. Barabasi, H. Jeong, Z. Neda, E. Ravasz, A. Schubert, T. Vicsek,
Physica A {\bf 311}, 590-614 (2002).
\bibitem{jeong_paper}
H. Jeong, Z. Neda and A.-L. Barab\'asi, 
Europhys. Lett. {\bf 61}, 567 (2003).
\bibitem{simeon_paper}
 S. Warner,
Library Hi Tech, {\bf 21}, No. 2, 151-158 (2003).
%
%
%
\bibitem{watts_actors}
D. J. Watts, S. H. Strogatz,
Nature {\bf 393}, 440 (1998).
\bibitem{barab_paper} 
A.-L. Barab\'asi, R. Albert
Science {\bf 286}, 509 (1999).
%
%
\bibitem{Kertesz}
 J.-P. Onnela, A. Chakraborti, K. Kaski, J. Kert\'esz, A. Kanto,
 Physica Scripta {\bf T106}, 48 (2003).
%
\bibitem{ErdosRenyi} 
P. Erd\H{o}s and A. R\'enyi, 
Publ.\ of the Math.\ Inst.\ of the Hung.\ Acad.\ of Sci.\ 
{\bf 5}, 17-61 (1960).
%
%
\bibitem{database}
The data was downloaded from the following database,
 provided by Hawoong Jeong,\\
{\tt http://www.nd.edu/\~{}networks/database/\\ index.html}
\bibitem{simeon_data}
The co-authorship data was 
 kindly provided by Simeon Warner via private communications. 
\bibitem{asset_data}
The asset graph sequence was kindly provided by J.-P. Onnela and 
J. Kert\'esz in private communications.
%
%
\bibitem{newman_pref}
M. E. J. Newman,
Phys. Rev. E {\bf 64}, 025102 (2001).
\bibitem{redner_pref}
P. L. Krapivsky, S. Redner, and F. Leyvraz,
Phys. Rev. Lett. {\bf 85}, 4629 (2000).
\end{thebibliography}
\end{document}